\documentclass[dvips,12pt]{article}
\usepackage{graphicx}
\newcommand{\doublespacing}{\let\CS=\@currsize\renewcommand{\baselinesstrech}
{2.0}\tiny\CS}

\begin{document}

\newcommand{\bc}{\begin{center}}
\newcommand{\ec}{\end{center}}
\newcommand{\vs}{\vspace}

\bc {\huge \bf Continuum States in } \ec

\bc {\huge \bf Generalized Swanson Models } \ec

\vs{.5cm}

\vspace{.5cm}

\begin{center}
{\it \large A. Sinha{\footnote {e-mail : anjana$_-$t@isical.ac.in, anjana23@rediffmail.com}}} \\
\end{center}

\begin{center}
{and}
\end{center}

\begin{center}
{\it \large P. Roy{\footnote {e-mail : pinaki@isical.ac.in}}} \\
\end{center}

\begin{center}
{\it \large Physics \& Applied Mathematics Unit \\
Indian Statistical Institute \\
Kolkata - 700 108 \\ INDIA }
\end{center}

\vspace{.5cm}

\begin{abstract}

A one-to-one correspondence is known to exist between the spectra
of the discrete states of the non Hermitian Swanson-type
Hamiltonian $ H = {\cal{A}}^{\dagger} {\cal{A}} + \alpha {\cal{A}}
^2 + \beta {\cal{A}}^{\dagger \ 2} $, ($\alpha \neq \beta $), and
an equivalent Hermitian Schr\"{o}dinger Hamiltonian $h$, the two
Hamiltonians being related through a similarity transformation. In
this work we consider the continuum states of $h$, and examine the
nature of the corresponding states of $H$.

\vspace{1cm}


\vspace{1cm}


\end{abstract}

\pagebreak


The spectrum of the non Hermitian Swanson Hamiltonian $ H =
a^{\dagger} a + \alpha a ^2 + \beta a^{\dagger \ 2} $ ($ \ \alpha
\neq \beta $), has been found to have a one-to-one correspondence
with that of the conventional (Hermitian) Harmonic oscillator
Hamiltonian $h$ \cite{swanson}. Subsequent works
\cite{quesne,jpa-07,jpa-08} have shown that by replacing the
Harmonic oscillator annihilation and creation operators $a$ and
$a^{\dagger}$ by generalized annihilation and creation operators
$A$ and $A^{\dagger}$ respectively, the Swanson Hamiltonian could
be generalized to include other interactions, and thus have the
generalized form $ H = {\cal{A}}^{\dagger} {\cal{A}} + \alpha
{\cal{A}} ^2 + \beta {\cal{A}}^{\dagger \ 2} $, ($ \ \alpha \neq
\beta $). In these cases too, $H$ can be mapped to an equivalent
Hermitian Hamiltonian $h$, with the help of a similarity
transformation (say $\rho $), and the bound state energies were
found to be the same for both the Hamiltonians (i.e., $H$ and
$h$). However, when $h$ has both discrete as well as continuous
spectra, no correspondence has been established till date between
the continuum states of $h$ and those (if any) of $H$. It may be
mentioned here that two works dealing with one-dimensional
scattering in non Hermitian quantum mechanics (and hence, dealing
with states having continuous spectra) deserve special mention in
this regard. The first of these is the review article on complex
absorbing potentials \cite{muga}, dealing with one dimensional
scattering in non-hermitian quantum mechanics, with particular
emphasis on complex ${\cal{PT}}$-symmetric potentials. The second
work gives a more detailed study of one-dimensional scattering in
${\cal{PT}}$-symmetric potentials in particular, with some
explicit examples of solvable potentials \cite{cannata1,cannata2}.

However, in this work we shall not investigate scattering in
generalized Swanson models. Rather, our aim is to focus our
attention on those states of $h$ having continuous energies, and
explore the nature of the corresponding states of $H$. We shall
also find the probability current density and charge density for
the $\eta$-pseudo Hermitian generalized Swanson Hamiltonian $H$.
Our studies will be based on two well known interactions viz., the
P\"{o}schl-Teller and Morse models.


\vspace{1cm}

It is well known by now that a quantum system described by a
$\eta$-pseudo Hermitian Hamiltonian $H$, can be mapped to an
equivalent system described by its corresponding Hermitian
counterpart $h$, with the help of a similarity transformation
$\rho$ \cite{jpa-07, jones,mostafa2},
\begin{equation}\label{similar}
    h = \rho H \rho ^{-1}
\end{equation}
So we start with the generalized Swanson Hamiltonian
\begin{equation}\label{genH}
    H = {\cal{A}}^{\dagger} {\cal{A}} + \alpha {\cal{A}} ^2 +
    \beta {\cal{A}}^{\dagger \ 2}  \ ,  \qquad \alpha \neq \beta
\end{equation}
with solutions $\psi (x)$ satisfying the eigen value equation
    $$ H \psi (x) = E \psi (x) $$
where $\alpha$ and $\beta$ are real, dimensionless constants.
Since we are dealing with non Hermitian $H$, hence $\alpha \ \neq
\ \beta $. Here $\cal{A}$ and ${\cal{A}}^{\dagger} $ are
generalized annihilation and creation operators given by
\begin{equation}\label{a}
    {\cal{A}} = \displaystyle \frac{1}{\sqrt{1 - \alpha - \beta}}
    \ \left\{ \frac{d}{dx} + W(x) \right\} \qquad ,
    \qquad {\cal{A}} ^{\dagger} = \displaystyle
    \frac{1}{\sqrt{1 - \alpha - \beta}} \ \left\{ - \frac{d}{dx}
    + W(x) \right\}
\end{equation}
If we apply to (\ref{genH}) a transformation of the form
\cite{faria}
\begin{equation}\label{transform}
    \psi (x) = \displaystyle \rho ^{-1}  \ \phi  (x)
\end{equation}
where
\begin{equation}\label{rho}
    \rho = \displaystyle e^{ - \mu \int W(x) ~ dx }
    \qquad , \qquad \rm{with} \ \ \ \mu = \displaystyle
    \frac{\alpha - \beta} {1 - \alpha - \beta}
    \ , \qquad  \alpha + \beta ~ \neq ~ 1
\end{equation}
(\ref{genH}) reduces to the following Hermitian
Schr\"{o}dinger-type Hamiltonian, with the same eigenvalue $E$ (in
units $\hbar = 2m = 1$)
\begin{equation}\label{h}
    h \ \phi (x) ~=~ \displaystyle \left( - \frac{d^2}{dx ^2} \ + \ V(x)
    \right) \phi (x) ~=~ E \phi (x)
\end{equation}
It may be recalled that the parameters $\alpha$ , $\beta$ must
obey certain constraints, viz., \cite{jpa-07,jpa-08}
\begin{equation}\label{constraint}
    \alpha + \beta < 1 \qquad \qquad , \qquad \qquad 4 \alpha
    \beta < 1
\end{equation}
Writing $V(x)$ in (\ref{h}) in the supersymmetric form $ ( w^2 -
w^{\prime} ) $ \cite{susy1,susy2}, $W(x)$ and $w(x)$ are found to
be inter-related through
\begin{equation}\label{ww}
    V(x) ~=~ w^2 (x) - w^{\prime} (x) ~=~ \displaystyle \left( \frac{
    \sqrt{1 - 4 \alpha \beta} }{ 1 - \alpha - \beta } \ W(x)
    \right) ^2  ~-~ \frac{1}{\left( 1 - \alpha -
    \beta \right)} W^{\ \prime} (x)
\end{equation}
Since we are dealing with $\eta$-pseudo Hermitian Hamiltonians,
$H$ obeys the relationship \cite{mostafa}
\begin{equation}\label{pseudo}
    H ^{\dagger} = \eta H \eta ^{-1} \qquad \qquad {\rm{or}}
    \qquad \qquad H ^{\dagger} \eta = \eta H
\end{equation}
where $\rho$ is the positive square root of $\eta$ \ ( $ \rho =
\sqrt{\eta} \ $ ) \cite{jpa-07,jones}. \\
So, whereas $\phi$ should obey the conventional normalization
conditions, $\psi$ should follow $\eta$ inner product and hence be
normalized according to $ \displaystyle < \psi _m \ | \ \eta \ | \
\psi _n > \ = \ \delta _{m,n} $ \cite{mostafa}. \\
Now let us look at the continuity condition for $\phi$, viz.,
\begin{equation}\label{continuity-phi}
    \displaystyle \frac{\partial}{\partial t} \phi ^* \phi
    + \nabla . j = 0
\end{equation}
where $ \phi ^* \phi $ represents the conventional charge density
and $j$ the conventional current density, given by
\begin{equation}\label{phi-j}
    j = \displaystyle i \left( \frac{\partial \phi ^*}{\partial x}
    \phi - \phi ^* \frac{\partial \phi }{\partial x} \right)
\end{equation}
in 1-dimension. \\
If we apply the inverse transformation of (\ref{transform}) to $
\phi $, then the equivalent continuity equation for $ \psi $ in
the pseudo Hermitian picture can be cast in the form
\begin{equation}\label{continuity-psi}
    \displaystyle \frac{\partial}{\partial t} \chi
    + \nabla . \bar{j} = 0
\end{equation}
provided we identify $\chi$ and $\bar{j}$ with
\begin{equation}\label{conti}
    \chi = \psi ^* \ \eta \ \psi \qquad , \qquad
    \bar{j} = \displaystyle i \eta \left( \frac{\partial
    \psi ^*}{\partial x} \psi -
    \psi ^* \frac{\partial \psi }{\partial x} \right)
\end{equation}
Thus $\chi$ plays the role of charge density and $ \bar{j} $ that
of current density for the generalized, pseudo Hermitian Swanson
Hamiltonian. The states of $H$ with discrete energies have been
found to have a one-to-one correspondence with similar states of
$h$. Our aim in this work is to see what happens to those states
of the Hermitian Hamiltonian $h$ with continuous spectra when we
go over to the corresponding non Hermitian Swanson Hamiltonian
$H$, with the help of the similarity transformation $ h = \rho H
\rho ^{-1} $. In particular, we shall do so for a couple of
explicit examples, viz., the P\"{o}schl-Teller and the Morse
models.

\vspace{1cm}

\noindent {\bf P\"{o}schl-Teller interaction :}

\vspace{.3cm}

Following \cite{jpa-08}, to map the non Hermitian Hamiltonian $H$
to its Hermitian Schr\"{o}dinger equivalent $h$, one needs to
solve the highly non trivial Ricatti equation (\ref{ww})
analytically. This is possible only if we take $W(x)$ and $w(x)$
to be of the same form. For the P\"{o}schl-Teller model, taking
\begin{equation}\label{w-PT}
    W(x) = \lambda_2 \sigma \ \tanh \ \sigma x
    \qquad , \qquad
    w(x) = \lambda_1 \sigma \ \tanh \ \sigma x
    \qquad \lambda _{1,2} > 0
\end{equation}
the potential term in the Schr\"{o}dinger-type Hermitian
Hamiltonian $h$ is obtained as
\begin{equation}\label{v-w}
    V(x) = \displaystyle \lambda _1 ^2 \sigma ^2 - \lambda _1 \left(
    \lambda _1  + 1  \right)\sigma ^2 \
    {\rm{sech}} ^2 \ \sigma x =
    \displaystyle \lambda _2 ^2 \sigma^2 \ \frac{1-4 \alpha \beta }{ ( 1 - \alpha -
    \beta )^2} \ - \ \zeta \sigma ^2 \ {\rm{sech}} ^2 \ \sigma x
\end{equation}
solving which gives the unknown parameter $\lambda _1$ in terms of
the known one $\lambda _2$ :
\begin{equation}\label{lambda}
    \lambda _1 = \frac{\sqrt{1 + 4 \zeta } - 1 }{2}
    \qquad {\rm{where}} \qquad
    \zeta = \frac{\lambda _2 ^2 (1-4 \alpha
    \beta) \ + \ \lambda _2  ( 1 - \alpha - \beta )}{ ( 1 - \alpha -
    \beta )^2}
\end{equation}
Introducing a new variable
\begin{equation}\label{y}
    y= \displaystyle  \cosh ^2 \sigma x
\end{equation}
and writing the solutions of $h$ as
\begin{equation}\label{phi-u}
    \phi = \displaystyle y^{\frac{ \lambda _1 + 1 }{2}} \ u(y)
\end{equation}
reduces equation (\ref{h}) to the hypergeometric equation
\begin{equation}\label{y-u}
    \displaystyle y(1-y) u^{ \ \prime \prime } + \left\{ \left(
    \lambda _1 + \frac{3}{2} \right) - \left( \lambda _1 + 2
    \right) y \right\} u^{ \ \prime} - \frac{1}{4} \left\{ \left(
    \lambda _1 + 1 \right) ^2 - \frac{\epsilon}{\sigma ^2}
    \right\} u = 0
\end{equation}
with complete solution
\begin{equation}\label{u-hypergeometric}
    u = \displaystyle A_1 \ _2F_1 \left( a,b,\frac{1}{2}; 1-y
    \right) + \displaystyle A_2 \ (1-y)^{1/2} \
    _2F_1 \left( a+\frac{1}{2},b + \frac{1}{2},\frac{3}{2}; 1-y \right)
\end{equation}
where \begin{equation}\label{epsilon}
    \epsilon = E - \lambda _1 ^2 \sigma ^2
\end{equation}
and $a,b$ are given below. Therefore, for  $\epsilon < 0$, say $
\epsilon = - \kappa ^2 \sigma ^2 $, the complete solution of $h$
is obtained as \cite{flugge}
\begin{equation}\label{boundstate-PT}
    \phi = \displaystyle A_1 \ y^{\frac{1}{2} (\lambda _1 +1 )}
    \ _2F_1 \left( a,b,\frac{1}{2}; 1-y
    \right) + \displaystyle A_2 \ y^{\frac{1}{2} (\lambda _1 +1 )}
    (1-y)^{1/2} \
    _2F_1 \left( a+\frac{1}{2},b + \frac{1}{2},\frac{3}{2}; 1-y \right)
\end{equation}
with
\begin{equation}\label{bound-ab}
    a = \displaystyle \frac{1}{2} \left( \lambda _1 + 1 -
    \frac{\kappa}{\sigma} \right)
    \qquad , \qquad
    b = \displaystyle \frac{1}{2} \left( \lambda _1 + 1 +
    \frac{\kappa}{\sigma} \right)
\end{equation}
Thus for negative energies, bound state normalizable solutions
exist only if either $a = -n$ or $ \displaystyle a + \frac{1}{2} =
-n$, giving
\begin{equation}\label{bounde-PT}
    E _n = - (\lambda _1 - n )^2 \sigma ^2 \qquad ,
    \qquad n = 0,1, \cdots \leq \lambda _1
\end{equation}
Consequently, the bound state energies $E$ for the generalized
Swanson Hamiltonian $H$, for this particular model, are also given
by (\ref{bounde-PT}). Since the form of $W(x)$ in (\ref{w-PT})
yields
\begin{equation}\label{rho-PT}
    \rho ^{-1} = \displaystyle ( \cosh \ \sigma x ) ^{\lambda _2
    \mu}
\end{equation}
the bound state solutions of $H$, obtained from
(\ref{boundstate-PT}) by applying (\ref{transform}), are found to
be
\begin{equation}\label{H-boundstate}
\begin{array}{lcl}
    \psi (x) &=& \displaystyle A_1 \ \cosh ^{\lambda _2 \mu + \lambda _1
    + 1 } \sigma x
    \ \ _2F_1 \left( a,b,\frac{1}{2}; - \sinh ^2 \sigma x
    \right) \\ \\
    & & + \ \displaystyle A_2 \ \cosh ^{\lambda _2 \mu + \lambda _1
    + 1 } \sigma x \
    \sinh \sigma x \ \
    _2F_1 \left( a+\frac{1}{2},b + \frac{1}{2},\frac{3}{2};
    - \sinh ^2 \sigma x \right) \\
\end{array}
\end{equation}
with either $A_1 = 0 $ or $ A_2 = 0$, for well behaved,
normalizable solutions; i.e., the eigenstates are either even
($A_2 = 0$)  or odd ($A_1 = 0$). However, in this work,  our
interest lies in those states of $H$ which correspond to the
continuum states (i.e., positive energy states, with $ \epsilon =
k^2 \sigma ^2 $.) of $h$, rather than the bound states. If $ \phi
(x) $ be the continuum states of $h$, with
\cite{cannata2,flugge,khare-jpa} :
\begin{equation}\label{phi-PT}
    \phi (x) = a_1 \phi _1 (x) + a_2 \phi _2 (x)
\end{equation}
then $\phi _1$ and $\phi _2$ are of the form
\begin{equation}\label{phi-1}
    \phi _1 = \displaystyle ( \cosh \  \sigma x)^{\lambda _1+1} \ \ _2 F _1
    \left( a,b,\frac{1}{2} ; - \sinh ^2 \sigma x \right)
\end{equation}
\begin{equation}\label{phi-2}
    \phi _2 = \displaystyle (\cosh \ \sigma x)^{\lambda _1+1} \ \sinh \
    \sigma x \ \ _2 F _1
    \left( a+ \frac{1}{2},b + \frac{1}{2},\frac{3}{2} ;
    - \sinh ^2 \sigma x \right)
\end{equation}
where
\begin{equation}\label{ab}
    a = \displaystyle \frac{1}{2} \left( \lambda _1+1+ i \frac{k}{\sigma} \right)
    \qquad , \qquad
    b = \displaystyle \frac{1}{2} \left( \lambda _1+1- i \frac{k}{\sigma} \right)
\end{equation}
The continuum states of $H$ are obtained (by applying eq
(\ref{transform}) to $\phi (x)$) in the same form as
(\ref{H-boundstate}), with $a,b$ as defined in (\ref{ab}). Using
the asymptotic limit of the Hypergeometric Functions $\ _2 F_1
(a,b,c;z)$ for large $\displaystyle \mid z \mid $ \cite{handbook},
viz.,
\begin{equation}\label{asymp-F}
\begin{array}{lcl}
    \displaystyle  _2 F_1 \left ( a,b,c;z \right) & \sim &
    \displaystyle
    \frac{\Gamma \left( c \right) \ \Gamma \left( b-a \right)
    }{\Gamma \left( b \right) \ \Gamma \left( c-a \right) }
    \ \left( -z \right) ^{-a} \ \displaystyle
    _2 F_1 \left( a,1-c+a,1-b+a; \frac{1}{z} \right) \\ \\
    \displaystyle
    & & \displaystyle + \ \frac{\Gamma \left( c \right) \ \Gamma \left( a-b \right)
    }{\Gamma \left( a \right) \ \Gamma \left( c-b \right) }
    \ \left( -z \right) ^{-b} \ \displaystyle
    _2 F_1 \left( b,1-c+b,1-a+b;\frac{1}{z} \right) \\ \\
\end{array}
\end{equation}
along with the fact that when $x \rightarrow \pm \infty $, $
\displaystyle - \ \sinh ^2 \ \sigma x \rightarrow - 2^{-2} \ e^{ 2
\sigma |x|} $, and $ \displaystyle \cosh ^2 \ \sigma x \rightarrow
2^{-2} \ e^{ 2 \sigma |x|} $, it can be shown by simple
straightforward algebra that the solutions $\phi$ have asymptotic
behaviour of the form \cite{flugge}
\begin{equation}\label{asymp-PT}
    \phi (x) = \left\{
    \begin{array}{lcl}
    \displaystyle  e^{ikx} + R e^{-ikx} \qquad , \qquad {\rm{for}} \ x < 0 \\
    \\
    \displaystyle  T e^{ikx} \qquad \qquad \ \ \ \ , \qquad {\rm{for}}
    \ x > 0
    \\
    \end{array}
    \right.
\end{equation}
provided the coefficients $a_1 , a_2$ obey certain restrictions,
giving the reflection and transmission amplitudes in
(\ref{asymp-PT}) by :
\begin{equation}\label{RT-PT}
    R = \displaystyle \frac{1}{2} \left( e^{2i \varphi _e } + e^{2i \varphi _o }
    \right)
    \qquad , \qquad
    T = \displaystyle \frac{1}{2} \left( e^{2i \varphi _e } - e^{2i \varphi _o }
    \right)
\end{equation}
with
\begin{equation}\label{varphi-e}
    \varphi _e = \displaystyle arg \frac{\Gamma
    (ik/\alpha) \ e^{-i\frac{k}{\alpha} \log 2}}{
    \Gamma \left( \frac{\lambda + 1}{2} + i \frac{k}{2 \alpha}
    \right) \
    \Gamma \left( \frac{- \lambda}{2} + i \frac{k}{2 \alpha}
    \right)}
\end{equation}
\begin{equation}\label{varphi-o}
    \varphi _o = \displaystyle arg \frac{\Gamma
    (ik/\alpha) \ e^{-i\frac{k}{\alpha} \log 2}}{
    \Gamma \left( \frac{\lambda }{2} + i \frac{k}{2 \alpha}
    \right) \
    \Gamma \left( \frac{1 - \lambda}{2} + i \frac{k}{2 \alpha}
    \right) }
\end{equation}
Thus the conventional conservation law $ |R|^2 + |T|^2 =
1 $ is obeyed. \\
We shall now use the relationship $ \psi (x) = \displaystyle \rho
^{-1}  \ \phi (x)$, to obtain those states of $H$ which correspond
to the continuum states of $h$, and hence possess continuous
energies. If these solutions are written as
\begin{equation}
    \psi (x) = B_1 \psi _1 (x) + B_2 \psi _2 (x)
\end{equation}
then, applying (\ref{transform}) to (\ref{asymp-PT}) yields the
following asymptotic behaviour of $\psi$ :
\begin{equation}\label{asympsi-PT}
    \psi (x) \sim \left\{
    \begin{array}{lcl}
    \displaystyle  e^{\lambda _2 \mu |x|}
    \left\{ e^{ikx} + R e^{-ikx} \right\} \qquad , \qquad {\rm{for}} \ x < 0 \\
    \\
    \displaystyle  e^{\lambda _2 \mu |x|}
    \ T e^{ikx} \qquad \qquad \qquad  \ \ , \qquad {\rm{for}} \ x > 0
    \\
    \end{array}
    \right.
\end{equation}
with $R$ and $T$ as defined in (\ref{RT-PT}). However, unlike in
the Hermitian case, here $R$ and $T$ can no longer be identified
as the reflection and transmission amplitudes. Hence the plane
wave solutions of the Hermitian Hamiltonian $h$ are replaced in
case of the pseudo Hermitian Swanson type Hamiltonian $H$, by
progressive waves for $ \mu > 0 $, i.e., $ \alpha > \beta $, or
damped waves for $ \mu < 0$, i.e., $ \alpha < \beta $. Thus the
role played by $\mu$, i.e. the parameters $\alpha , \beta$ is
quite significant in case of the continuum states. However, for
the bound states no significant change is introduced in the nature
of the solutions of $h$ and $H$, depending on the sign of $\mu$.
For instance, for $ \alpha = \displaystyle \frac{1}{8} \ , \ \beta
= \displaystyle \frac{1}{4} \ , \ \lambda _2 = 5 $, we obtain $
\zeta = 384 \ , \ \mu = \displaystyle - \frac{1}{5} \ , \ \lambda
_1 = 9.31 $, implying that there are ten bound states ($ n = 0 , 1
, 2 , \cdots , 9 $) in both $h$ as well as $H$ \cite{jpa-07}. The
density function $\psi ^* \psi = \tau (x) $ (say) for such a
damped wave solution for $ \psi $ as given in (\ref{asympsi-PT})
is plotted in Fig. 1, for the parameter values mentioned above,
i.e., $ \alpha = \displaystyle \frac{1}{8} \ , \ \beta =
\displaystyle \frac{1}{4} \ , \ \lambda _2 = 5 , \ \mu = -1/5 \ $,
while Fig. 2 shows the density function $ \tau (x) $ for a
progressive wave solution for the same $ \psi $ for the parameter
values $ \alpha = \displaystyle \frac{1}{2} \ , \ \beta =
\displaystyle \frac{1}{4} \ , \ \lambda _2 = 1 , \ \mu = 1 \ , $
so that $ \zeta = 12 $ and $ \lambda _1 = 3 $.


{\begin{figure}[hp]
\begin{center}
\scalebox{0.8}{\includegraphics{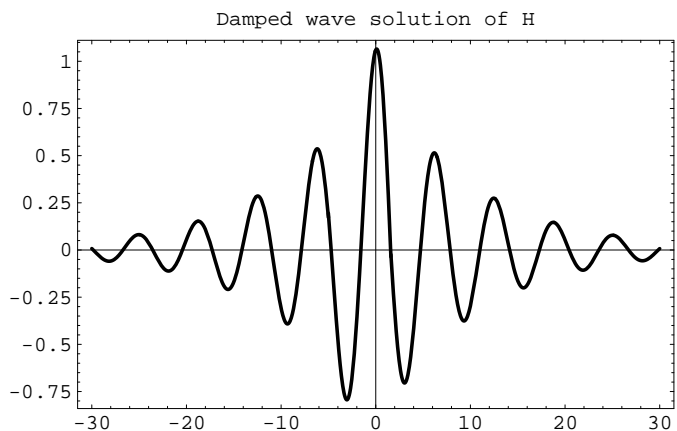}} \label*{}\caption{\small
{A plot of Re $ \psi $ vs $x$ for the damped wave solutions with
continuous spectra, for $-$ve $\mu$ }}
\end{center}
\end{figure}}


\pagebreak

{\begin{figure}[hp]
\begin{center}
\scalebox{0.8}{\includegraphics{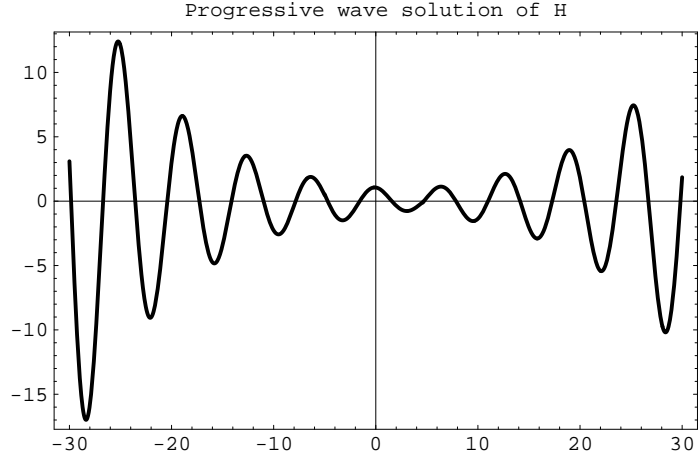}} \label*{}\caption{\small
{A plot of Re $ \psi $ vs $x$ for the progressive solutions with
continuous spectra, for $+$ve $\mu$ }}
\end{center}
\end{figure}}


\vspace{1cm}

\noindent {\bf Morse interaction :}

\vspace{.3cm}

To check whether the results obtained so far are peculiar to the
particular interaction studied, we consider a second example,
viz., the Morse model. Taking the following ansatz for $W(x)$ and
$w(x)$ :
\begin{equation}\label{w-Morse}
    W(x) = \displaystyle a_2 \sigma - b_2 \sigma \ e^{ - \sigma x}
    \qquad , \qquad
    w(x) = \displaystyle a_1 \sigma - b_1 \sigma \ e^{ - \sigma x}
    \qquad a_{1,2} , b_{1,2} > 0
\end{equation}
This yields the following expression for $V(x)$ in equation
(\ref{ww}) :
\begin{equation}\label{ww-morse}
\begin{array}{lll}
    V(x) & = & \ \displaystyle a_1 ^2 \sigma ^2 + b_1 ^2 \sigma ^2 e^{- 2
    \sigma x} - 2 b_1 \left( a_1 + \frac{1}{2} \right) \sigma ^2 e^{-
    \sigma x } \\ \\
    & = & \displaystyle \frac{1 - 4 \alpha \beta }{ \left( 1 -
    \alpha - \beta \right) ^2} a_2 ^2 \sigma ^2 +
    \frac{1 - 4 \alpha \beta }{ \left( 1 -
    \alpha - \beta \right) ^2}  b_2 ^2 \sigma ^2 e^{- 2
    \sigma x} \\ \\
    & & \displaystyle - \frac{(1 - 4 \alpha \beta) 2 a_2 +
    (1 - \alpha - \beta ) }{(1 - \alpha - \beta )^2}
    \ b_2 ^2 \sigma ^2 e^{- \sigma x} \\
\end{array}
\end{equation}
with
\begin{equation}\label{a1b1}
    a_1 = \displaystyle \frac{\sqrt{1 - 4 \alpha \beta}}{1 -
    \alpha - \beta } \ a_2 \ + \ \frac{1}{2 \sqrt{1 - 4 \alpha
    \beta}}\ - \frac{1}{2}
    \qquad , \qquad
    b_1 = \displaystyle \frac{\sqrt{ 1 - 4 \alpha \beta }}{ 1 -
    \alpha - \beta } \ b_2
\end{equation}
Substituting
\begin{equation}\label{z}
    z = \displaystyle 2 b_1 e^{ - \sigma x} \qquad , \qquad
    \phi  = \displaystyle e^{ - z/2} z^s u(z)
\end{equation}
so that $ - \infty < x < \infty $ transforms over to $ 0 \leq z <
\infty $, and eq. (\ref{h}) reduces to
\begin{equation}\label{u}
    \displaystyle u^{\ \prime \prime} + \left( \frac{2s+1}{z} - 1
    \right) u^{\ \prime} + \left( \frac{a_1 - s }{z}+
    \frac{\epsilon / \sigma ^2 + s^2}{z^2} \right) u = 0
\end{equation}
where $ \epsilon = E - a_1 ^2 \sigma ^2 $. For bound states, $
\epsilon = - \kappa ^2 $ giving $ s = \pm \kappa / \sigma $. Thus
the bound state solutions of the Hermitian Hamiltonian $h$ and its
corresponding non Hermitian one $H$ are respectively given by
\cite{susy1},
\begin{equation}\label{phi-morse}
    \phi _n = \displaystyle e^{-z/2} z^s L_n ^{2s} (z)
\end{equation}
\begin{equation}\label{psi-morse}
    \psi _n = \rho ^{-1} \phi _n
    \sim \displaystyle e^{- \left( 1 + \mu b_2 / b_1 \right) z/2}
    z^{ s - \mu a_2 }  L_n ^{2s} (z)
\end{equation}
where $ L_n ^{2s} $ are associated Laguerre polynomials
\cite{handbook}, $ s = a_1 - n $ and normalization requirement
restricts $s$ to positive values only. The bound state energies of
both the Hermitian as well as non Hermitian systems are obtained
as
\begin{equation}\label{e-morse}
    \epsilon _n = \displaystyle - \left( a_1 - n \right) ^2 \sigma
    ^2 \qquad {\rm{giving}} \qquad E_n = \displaystyle
    \left( 2 a_1  - n \right) n \sigma ^2  \qquad , \qquad n = 0,
    1, 2, \cdots < a_1
\end{equation}
However, our aim in this work is to explore those states of $H$
whose Hermitian equivalents are the continuum states (i.e.
positive energies) of $h$. We proceed in a manner analogous to the
previous section, with the following form of $\rho$ for this model
\begin{equation}\label{rho-morse}
    \rho = \displaystyle \left( \frac{z}{2b_1} \right) ^{\mu a_2}
    e^{ - \frac{ \mu b_2 }{2b_1} z} \qquad , \qquad
    {\rm{where}} \qquad
    \displaystyle \frac{ \mu b_2 }{2b_1} = \frac{\alpha - \beta }{
    2 \sqrt{1 - 4 \alpha \beta}}
\end{equation}
Writing $ \epsilon = E - a_2 \sigma ^2 = k ^2 $, so that $ s =
\displaystyle \pm \frac{ik}{ \sigma} $, the continuum state
solutions of $h$ are found to be \cite{alhassid}
\begin{equation}\label{phi-scattmorse}
    \phi = \displaystyle A_1 \ e^{-z/2} \ z^{ik / \sigma}
    \ F \left( \frac{ik}{\sigma} - a_1 , \frac{2ik}{\sigma} + 1 ,
    z \right) + A_2 \ e^{-z/2} \ z^{- ik / \sigma}
    \ F \left( - \frac{ ik}{\sigma} - a_1 , - \frac{2ik}{\sigma} + 1 ,
    z \right)
\end{equation}
where $F(a,b,z)$ are the Kummer confluent Hypergeometric functions
\cite{handbook}. Since the potential $V(x)$ goes to infinity for
large negative values of $x$ (i.e., $z \rightarrow 0$), the wave
function $\phi (x)$ should vanish in this region : $ \phi (x
\rightarrow - \infty) \rightarrow 0 $. Using the properties of
confluent Hypergeometric functions \cite{handbook},
\begin{equation}\label{phi-0}
    \phi ( z \rightarrow 0) \rightarrow \displaystyle A_1 \frac{
    \Gamma \left( \frac{2ik}{\sigma} + 1 \right) }{
    \Gamma \left( \frac{ik}{\sigma} - a_1 \right)}
    \ + \
    A_2 \frac{
    \Gamma \left( - \frac{2ik}{\sigma} + 1 \right) }{
    \Gamma \left( - \frac{ik}{\sigma} - a_1 \right)}
    \ \rightarrow \ 0
\end{equation}
giving the ratio $ A_2 : A_1 $ as
\begin{equation}\label{a12}
    \displaystyle \frac{A_2}{A_1}  = \displaystyle - \ \frac{
    \Gamma \left( \frac{2ik}{\sigma} + 1 \right)
    \Gamma \left( - \frac{ik}{\sigma} - a_1 \right)}{
    \Gamma \left( - \frac{2ik}{\sigma} + 1 \right)
    \Gamma \left( \frac{ik}{\sigma} - a_1 \right)}
\end{equation}
Similarly for large values of $x$ in the positive direction,
\begin{equation}\label{phi-infty}
    \phi ( x \rightarrow \infty) \rightarrow \displaystyle A_1
    (2b_1 )^{ik / \sigma} e^{-ikx}
    \ + \
    A_2 (2b_1 )^{- ik / \sigma} e^{ikx}
\end{equation}
Thus $\phi (x)$ can be written in the form (\ref{RT-PT}), with
\begin{equation}\label{r-morse}
    R = \displaystyle (2b_1 )^{- ik / \sigma} \ \frac{
    \Gamma \left( \frac{2ik}{\sigma} + 1 \right)
    \Gamma \left( - \frac{ik}{\sigma} - a_1 \right)}{
    \Gamma \left( - \frac{2ik}{\sigma} + 1 \right)
    \Gamma \left( \frac{ik}{\sigma} - a_1 \right)}
\end{equation}
and $T = 0$, which is expected as the potential goes to $ \infty $
at $ x \rightarrow - \infty $. Instead of going into detailed
calculations, we just quote the solutions of $H$ having a
one-to-one correspondence with those states of $h$ having
continuous spectra :
\begin{equation}\label{psi-morse}
\begin{array}{lcl}
    \psi \ (x \rightarrow \infty)
    & \sim & \displaystyle \left( \frac{z}{2b_1} \right) ^{- \mu
    a_2 } \ e^{ \frac{\mu b_2 }{2b_1} z} \phi \\ \\
    & = & \displaystyle e^{ \mu a_2 \sigma x} \left( c_1 e^{- ikx}
    \ + \ c_2 e^{ikx} \right) \\
\end{array}
\end{equation}
Once again, we obtain a result identical to that obtained in the
previous case viz., depending on the sign of $ \mu $, states with
continuous spectra are either progressive waves ( $ \mu > 0 $) or
damped waves ( $ \mu < 0 $) in the non Hermitian generalized
Swanson Hamiltonian.


\vspace{1cm}

To conclude, we have studied states with continuous spectra in a
class of $\eta$-pseudo Hermitian Hamiltonians, which are of the
generalized Swanson type, viz., $H =
    {\cal{A}}^{\dagger} {\cal{A}} + \alpha {\cal{A}} ^2 +
    \beta {\cal{A}}^{\dagger \ 2}  \ ,  \ \ \alpha \neq \beta$,
and hence can be mapped to an equivalent Hermitian
Schr\"{o}dinger-type Hamiltonian $h$ with the help of a similarity
transformation $ \rho$. We have also obtained the modified
continuity equation the solutions of such non Hermitian
Hamiltonians should obey, and obtained new definitions for the
charge density and current density. In particular, we have
analyzed the positive energy solutions for two one-dimensional
pseudo Hermitian models, whose Hermitian equivalents are the
P\"{o}schl-Teller and the Morse interactions. In both the cases it
is observed that the relative strength of the parameters $ \alpha
$ and $ \beta $ plays a crucial role in determining the nature of
the solutions with continuous spectra. If $ \alpha > \beta $
implying $ \mu > 0 $, then the continuum states of $h$ represented
by plane wave solutions are replaced by progressive waves in case
of the corresponding non Hermitian Hamiltonian $H$. Similarly, for
$ \alpha < \beta $ implying $ \mu < 0 $, the continuum energy
states of the generalized Swanson Hamiltonian $H$ are represented
by damped waves. This asymmetrical nature of the parameters $
\alpha , \beta $ is quite interesting. For $\mu < 0$, since the
solutions $\psi (x) $ are damped at $ \pm \infty $, the density
function for these states rapidly goes to zero as $|x|$ increases.
Thus these solutions may be interpreted as bound states in the
continuum. In other words, the scattering states of the hermitian
problem get transformed to bound states in the continuum for the
corresponding non Hermitian problem. It is worth noting here that
with modified definitions for current density and charge density
for non Hermitian generalized Swanson models, there is no
violation in the equation of continuity  in either case.

We have plotted the density function, viz., $ \tau = \psi ^* \psi
$ for the damped wave solution ($ \mu < 0 $) for the
P\"{o}schl-Teller model in Fig. 1, while the density function for
a progressive solution ($ \mu < 0 $) for the same model is plotted
in Fig. 2. The Morse interaction shows similar behaviour.

\vspace{1cm}

\section{Acknowledgment}

The authors would like to thank the referee for instructive
criticism. This work was partly supported by SERC, DST, Govt. of
India, through the Fast Track Scheme for Young Scientists
(SR/FTP/PS-07/2004), to one of the authors (AS).

\end{document}